\documentclass[aps,pre,twocolumn,nopacs]{revtex4-1}
\usepackage{graphicx}
\usepackage{amssymb,amsfonts,amsmath}
\usepackage[colorlinks=true,citecolor=Cerulean,linkcolor=RubineRed,urlcolor=Cerulean]{hyperref}
\hypersetup{breaklinks=true}
\usepackage{graphicx}
\usepackage{color}
\usepackage[usenames,dvipsnames]{xcolor}
\usepackage{epstopdf}
\usepackage{units}

\usepackage{booktabs}
\usepackage{mathrsfs}
\usepackage{braket}
\usepackage{dsfont}
\usepackage{float}

\newcommand{\Heff}{H_\mathrm{eff}}

\newcommand{\aver}{\langle r \rangle}

\begin{document}

\title{Spectral statistics of driven Bose-Hubbard models}

\author{Jes\'us Mateos}
\affiliation{Departamento de F\'isica de Materiales, Universidad
Complutense de Madrid, E-28040 Madrid, Spain}

\author{Fernando Sols}
\affiliation{Departamento de F\'isica de Materiales, Universidad
Complutense de Madrid, E-28040 Madrid, Spain}

\author{Charles Creffield}
\affiliation{Departamento de F\'isica de Materiales, Universidad
Complutense de Madrid, E-28040 Madrid, Spain}

\date{\today}

\begin{abstract}
We study the spectral statistics of a one-dimensional Bose-Hubbard
model subjected to kinetic driving; a form of Floquet engineering 
where the kinetic energy is periodically driven in time with a zero time-average.
As the amplitude of the driving is increased, the ground state of the 
resulting flat-band system passes from the Mott insulator regime to an exotic 
superfluid. We show that this transition is accompanied by a change in the 
system's spectral statistics from Poisson to GOE-type. Remarkably, and unlike 
in the conventional Bose-Hubbard model which we use as a benchmark, the details 
of the GOE statistics are sensitive to the parity
of both the particle number and the lattice sites. We show how this effect
arises from a hidden symmetry of the Hamiltonian produced
by this form of Floquet driving.
\end{abstract}

\maketitle

\section{Introduction}
In recent years Floquet engineering 
\cite{engineering_review,simonet_review} has become 
an increasingly important tool to control the dynamics of quantum systems.
In this approach, a system is driven by a time-periodic perturbation, allowing
its dynamics to be decomposed into two parts: a
so-called micromotion oscillating at the same frequency as the
driving, and an effective static Hamiltonian. In the limit of high frequency,
the micromotion can frequently be neglected, and as a result the driven
system can be simply described by just the effective Hamiltonian, the parameters
of which can be controlled by varying the amplitude or phase of the driving.
A typical example is controlling the tunneling of a particle in a lattice 
potential. By shaking the lattice periodically in time, 
the tunneling becomes renormalized \cite{dunlap}, allowing it
to be tuned \cite{lignier} by altering the parameters of the shaking.
In this way the effective tunneling can be set
to zero to produce the effect known as
coherent destruction of tunneling \cite{CDT}, 
to negative values to produce negative effective mass \cite{haller,solitons}, or can
be rendered complex \cite{jaksch_zoller,Creffield_2016}
to simulate the effect of an applied magnetic field \cite{ketterle,bloch}. 

\begin{figure}[htb]
\begin{center}
\includegraphics[width=0.45\textwidth,clip=true]{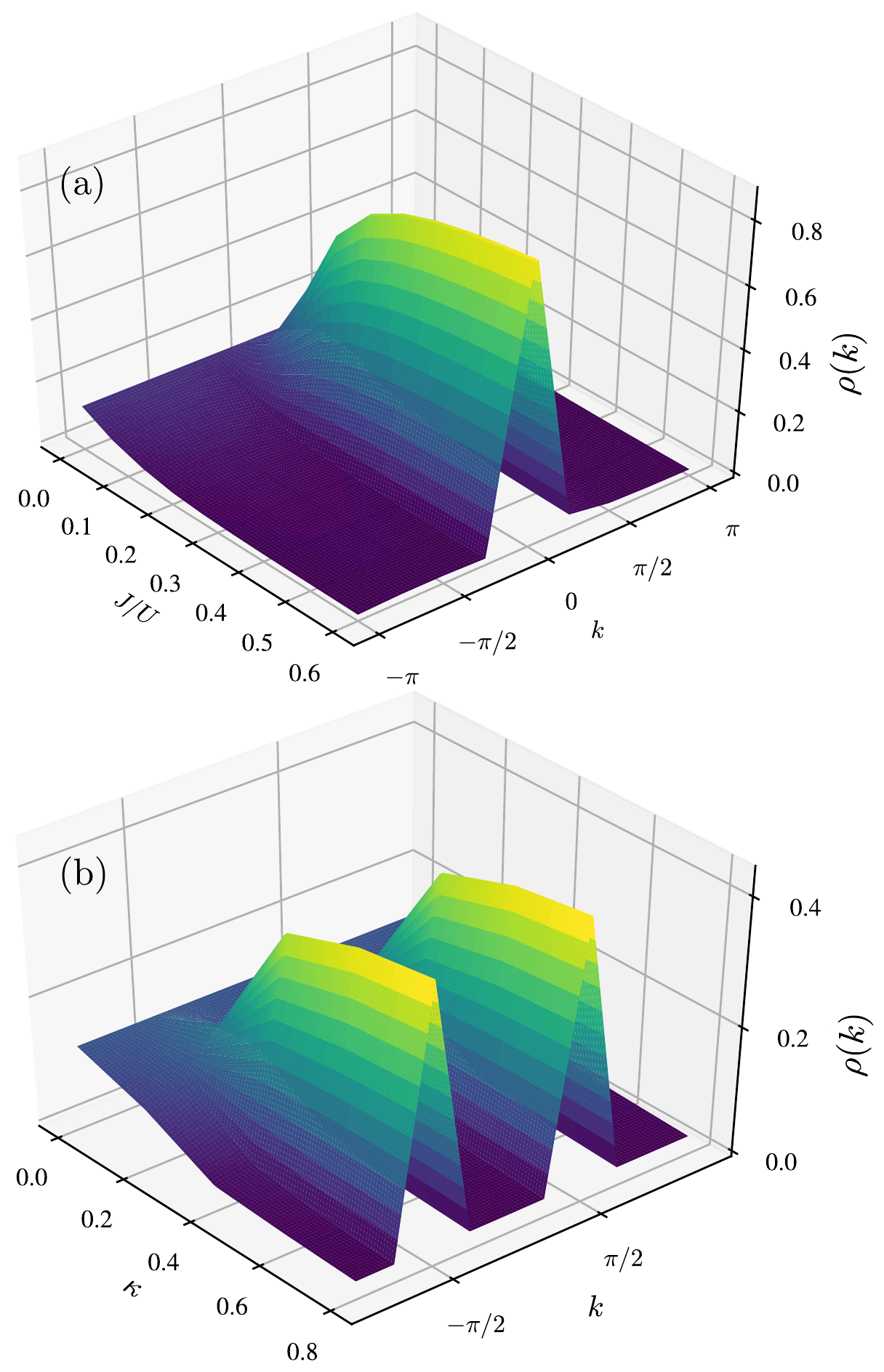}
\end{center}
\caption{
The single-particle momentum density of the ground state of the 
Bose-Hubbard model.
(a) For small $J/U$ the ground state of the conventional BH model is a Mott insulator, and the momentum
density is flat. As $J/U$ is increased the system becomes superfluid, and
a single peak develops at zero momentum. 
(b) Kinetic driving. At low values of $\kappa$ the system is again
a Mott insulator with a uniform momentum density. For larger $\kappa$ the system
makes a transition to an exotic superfluid, characterized by the occupation
of {\em two} momentum states, $k = \pm \pi/2$.
Parameters: 8 sites, 8 particles.}
\label{phase_transition}
\end{figure}

This type of fine control over the parameters of the effective Hamiltonian
allows the simulation \cite{maciej,cirac_zoller} of physical problems in which the
effective Hamiltonian is engineered to mimic the Hamiltonian of another
system \cite{rydberg_1,rydberg_2,rydberg_3}, and also to encode problems in pure mathematics \cite{factors,riemann}. 
The Floquet method is not only applicable at the single-particle level, but is also 
relevant to controlling  strongly 
correlated systems. Regulating the effective tunelling, for example,
can be used to coherently induce the Mott transition in the conventional
Bose-Hubbard model \cite{holthaus_2005,monteiro_2006}
(see Fig. \ref{phase_transition}a).
In many-body systems the periodic driving will eventually cause the system to 
heat up to infinite temperature \cite{heating_1,heating_2,heating_3}. 
It has been shown, however, that a generic system will first pass through
a long-lived ``pre-thermal'' regime, in which the rate of heating is
exponentially small in the driving frequency
\cite{suppression_1,suppression_2,suppression_3,suppression_4,romero}.
For high-frequency driving it is thus possible to manipulate the properties 
of a system using Floquet engineering  while it remains in this prethermal state. 
Although in quantum simulation applications one is principally interested
in the ground state of the effective Hamiltonian (such as whether it is
an insulator or a superfluid), it is important to note that Floquet
engineering affects the {\em entire} spectrum of the driven system,
and so also controls the statistical distribution
of the energy level spacings.

This distribution is frequently used to address
the question of whether a system is integrable or chaotic.
Berry and Tabor \cite{berry_tabor} conjectured that Poisson statistics
would describe systems having an integrable classical limit,
while conversely the Bohigas conjecture \cite{bohigas} indicates that systems
with a semiclassical chaotic limit would be described by random
matrix theory (RMT). Combining these results thus allows the spectral statistics
to test whether a given system is chaotic (RMT) or 
integrable (Poisson). 
Quantum many-body systems, which do not necessarily
possess a well-defined semiclassical limit, also seem to follow this
rule in general \cite{altshuler}, although some exceptions are
known \cite{kazue,benet_2003,armando_2004}. 
Depending on its symmetries, the spectrum of a quantum chaotic Hamiltonian
will therefore be expected to fall into one of the three ensembles of RMT.
In particular, if such a Hamiltonian possesses time-reversal symmetry, it 
should belong to the Gaussian orthogonal ensemble (GOE).
This has been verified for
a large number of many-body systems, including the Heisenberg,
$t-J$, and fermionic Hubbard models \cite{poilblanc_1993,hsu},
the Bose-Hubbard model \cite{kolovsky_2004,kollath_2010},
the ionic Hubbard model \cite{demarco_2022},
and the SYK model \cite{haque_2019}.

In this work we study the statistics of the Floquet spectrum
of the Bose-Hubbard model under kinetic driving \cite{njp_2018}, in 
which the hopping parameter is driven periodically in time
with a zero time-average. This results in the suppression of 
first-order single-particle hopping, which places the setup in the 
class of flat-band systems, generally characterized by a vanishing 
single-particle group velocity. Other possible origins of flat-band 
behavior are frustration \cite{F5}, spin-orbit coupling \cite{F7} and, 
importantly, the destructive interference between different paths in 
an elementary hopping process \cite{F6,F8,F9,F10,F11,F12}.

Just as
in the case of the conventional Bose-Hubbard model, the ground-state
of the system can be
tuned to pass from a Mott insulator state to a superfluid, as shown in
Fig. \ref{phase_transition}b, as the driving parameter is varied.
The resulting superfluid, however, is rather exotic due to
the unusual pairing correlations induced by the driving, 
and consists of a cat-like superposition of two many-body states of
opposite momentum \cite{prr_2019,njp_2023}. 
We will show that these unusual pairing effects not only manifest 
in the ground-state properties of the system, but also
in the spectral statistics of the model. Moreover, they are associated 
with a discrete symmetry non-existent in the undriven system. 
As a result, and
unlike the conventional Bose-Hubbard model, the
spectral statistics become sensitive to the parities of the number
of particles in the system and the number
of lattice sites.
When this symmetry is absent, the statistics is described by a single GOE, 
but by a mixture of two GOEs when it is present.

\section{Models}
The conventional Bose-Hubbard (BH) model is given by the Hamiltonian
\begin{equation}
H_{\mathrm{BH}} = -J \sum_{\langle i,j \rangle}
\left( a_i^\dagger a_j + H.c. \right) +
\frac{U}{2} \sum_j n_j \left( n_j - 1 \right) \ ,
\label{CBH}
\end{equation}
where $J$ is the hopping amplitude, $a_j / a_j^\dagger$
are bosonic annihilation/creation operators for a particle on site $j$, 
$n_j$ is the standard number operator $n_j = a^\dagger_j a_j$,
and $U$ is the Hubbard repulsion (that is, we take $U$ to be positive). 
The behavior of the model is dictated
by the competition between the kinetic energy, regulated by $J$,
and the interparticle repulsion $U$. 
For the case of commensurate filling, when the number of
bosons is an integer multiple of the number of sites,
the eigenstates of the system are approximately Fock states
in the limit $U \gg J$.
In particular the ground state consists of bosons that are highly localized
in space, forming a Mott insulator. 
As shown in Fig. \ref{phase_transition}a, the corresponding momentum density
is flat.  When $J/U$ is increased, the bosons are able to delocalise
and the phase-coherence of the state increases. As a result
a crossover from the Mott state to a superfluid 
\footnote{For ease of notation we use the term ``superfluid'', although in one 
dimension a system cannot have true long-range order. This, however, is not an
important limitation in realistic finite-size systems.}
occurs
at a critical coupling \cite{cazalilla} of $J / U_c \simeq 0.2$. This is
marked by a sharp peak forming at $k = 0$ in the momentum density, indicating 
the formation of a macroscopically-occupied zero momentum-state.
For convenience we shall henceforth label these two regimes by the character
of the ground state of the commensurate system, and term them the 
``Mott regime'' and the ``superfluid regime'' respectively. An additional regime
occurs in the limit $U \ll g$ when the system evolves towards a gas
of free bosons, which differs from a conventional superfluid
in certain physical aspects \cite{njp_2023}.

\subsection{Floquet theory}
A quantum system whose Hamiltonian varies
periodically with time, $H(t) = H(t + T)$ where $T$ is the
period of the system, can be described efficiently in terms
of Floquet theory. Such a time-dependent Hamiltonian can
arise, for example, if the system is driven by an oscillating
external field, or if a parameter of $H$ is periodically varied in time. 
We seek solutions of the time-dependent Schr\'odinger equation
$\left[ H(t) - i \hbar \partial_t \right] \phi_n(t) = \epsilon_n \phi_n(t)$.
The eigenfunctions $\phi_n(t)$ are $T$-periodic functions of time
termed Floquet states, while the eigenvalues $\epsilon_n$ are
generalizations of the energy eigenvalues obtained in static systems,
and are termed Floquet quasienergies. This type of solution is familiar
in the context of solid state physics, where {\em spatial} periodicity
allows spatial wavefunctions to be written in terms of Bloch states
and quasimomenta (Bloch's theorem). 
It should be noted that the quasienergies are only
defined up to integer multiples of the driving frequency $\omega$, and
so similarly to the case of quasimomentum, the quasienergy has a
Brillouin zone structure.

The Floquet states provide a complete basis, and so the time evolution
of a general quantum state can be expressed as
\begin{equation}
\psi(t) = \sum_n c_n e^{-i \epsilon_n t} \phi_n (t) \ ,
\label{expansion}
\end{equation}
analogous to the standard expansion of a wavefunction in energy eigenstates
of a static Hamiltonian. In the limit of high frequency,
we can see that a separation of energy
scales is present in this expression. The Floquet states have the
same periodicity as $H(t)$, and so in this limit 
frequencies they only produce structure over short timescales --
the so-called ``micromotion''. In contrast, behaviour on
timescales much longer than
$T$ are essentially determined by just the quasienergies.
In this limit we can thus consider the dynamics to be given simply
by an effective static Hamiltonian, $\Heff$, whose eigenvectors and
eigenvalues correspond to the Floquet states and quasienergies respectively.

It is usually difficult, however, to obtain analytical expressions
for this effective Hamiltonian. A common approach is to make
an expansion in inverse powers of the frequency
\begin{equation}
\Heff = \sum_{j=0}^\infty \frac{1}{\omega^j} \ \Heff^{(j)} \ ,
\label{magnus}
\end{equation}
such as the Magnus \cite{magnus} or van Vleck \cite{van_vleck} series.
If this expansion is well-defined and stable, then
one can obtain a high-frequency approximation to $\Heff$ by truncating
it to the lowest-order terms. This approximation will become
progressively more accurate as $\omega$ is increased.

\subsection{Potential driving}
To introduce the technique of Floquet engineering, we first consider
the well-studied case of when the Hamiltonian is modified by
a time-dependent potential.
By  periodically accelerating and decelerating the lattice in space, or
``shaking'' it, the Bose-Hubbard Hamiltonian (\ref{CBH}) seen in the rest frame of the
lattice acquires a time-dependent term, $H(t) =  H_{\mathrm{BH}} + V(t)$,
where the time-dependent potential is given by
\begin{equation}
V(t) = K \cos \omega t \sum_j j \ n_j \ ,
\end{equation}
$K$ being the amplitude of the shaking 
and $\omega$ is its frequency.
By transforming to the interaction picture and averaging over one
period of the driving, one obtains an effective Hamiltonian \cite{holthaus_2005}
valid in the limit of high driving frequencies ($\omega \gg U, J$)
{\em identical} to Eq. \ref{CBH}, but with a renormalized tunneling
parameter
$J \to J \mathcal{J}_0 \left(K / \omega \right)$, where ${\mathcal J}_0$
is the zeroth Bessel function of the first kind. 

The process of Floquet engineering then consists of adjusting the
parameters of the driving, $K$ and $\omega$, to regulate the amplitude
of the effective tunneling. In this way the value of $J / U$ can be tuned \cite{zenesini}
across the Mott-superfluid transition, without addressing the lattice
parameters directly. In particular, if the driving is tuned
to $K / \omega \simeq 2.404$ -- the first zero of the Bessel function --
the effective tunneling vanishes and the system becomes a Mott insulator.

\subsection{Kinetic driving}
In principle any term in (\ref{CBH})
can be periodically driven to obtain an effective Hamiltonian in the high-frequency
limit. Here we will consider the specific case of ``kinetic driving''
in which the tunelling parameter, $J$, is periodically oscillated with 
zero time-average
\begin{equation}
J(t) = J \cos \omega t \ .
\label{kinetc}
\end{equation}
A scheme for producing a driving of this type in experiment was
described in Ref. \cite{njp_2018}.

To obtain the effective Hamiltonian corresponding to this form of driving, it
is advantageous to work in a momentum representation. In momentum
space the conventional BH model (\ref{CBH}) can be written as
\begin{align}
        \nonumber
        H_{\mathrm{BH}}=-&2J\sum_{\ell=0}^{L-1}\cos(k_\ell)a^\dagger_{k_\ell}a_{k_\ell}\\
        \label{eq:Hcbh}
        +&\frac{U}{2L}\sum_{\ell, m, n, p=0}^{L-1} \delta_{k_\ell+k_m, k_n+k_p}
                             a^\dagger_{k_p} a^\dagger_{k_n}a_{k_m}a_{k_\ell},
\end{align}
where the momenta run over the first Brillouin zone (FBZ)
$k_\ell=2\pi\ell/L$, $\ell\in\mathds{Z}$ and $L$ is the number
of lattice sites. Performing the same
form of Floquet analysis as before \cite{njp_2018}
yields the effective Hamiltonian
\begin{align}
\nonumber
\Heff = & \frac{U}{2L}  \sum_{\ell, m, n, p=0}^{L-1}
\delta_{k_\ell+k_m, k_n+k_p}\\
\label{eq:Hkdbh}
& \times \mathcal{J}_{0}\left[2\kappa
F(k_\ell,k_m,k_n,k_p)\right] a^\dagger_{k_p}
a^\dagger_{k_n}a_{k_m}a_{k_\ell},
\end{align}
where $\kappa= J / \omega$ is the driving parameter, and
\begin{align}
\label{eq:Ffunction}
F(k_\ell,k_m,k_n,k_p)=\cos(k_\ell)+\cos(k_m)
-\cos(k_n)-\cos(k_p).
\end{align}
We will term this Hamiltonian the kinetically-driven BH model.
Note that since this Hamiltonian was derived in the high-frequency
limit, the driving parameter $\kappa$ is necessarily limited
to the range $\kappa < 1$.

\begin{figure}[htb]
\begin{center}
\includegraphics[width=0.6\textwidth,clip=true]{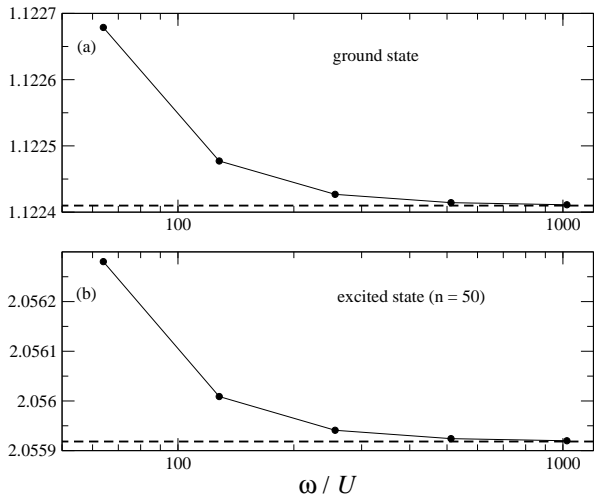}
\end{center}
\caption{
Representative quasienergies of the kinetically-driven BH model obtained from the exact diagonalization
of the full Floquet Hamiltonian, compared with the results of the
effective Hamiltonian (\ref{eq:Hkdbh}).
(a) Ground state. As the frequency, $\omega$ of the driving
is increased, the lowest quasienergy monotonically
approaches the ground state energy of the effective model, the difference
reducing as $\left( U / \omega \right)^2$.  
(b) As in (a) but for an arbitrarily chosen excited state ($n = 50$). As
before the quasienergy of the kinetically-driven BH model asymptotically approaches that
predicted by the effective model as $\left( U/\omega \right)^2$.
Clearly the effective model becomes exact in the limit $\omega \rightarrow \infty$.
Parameters: 8 sites, 8 particles, $\kappa = 0.3$. Energies measured in units of $J$.}
\label{energy_scaling}
\end{figure}

This effective Hamiltonian constitutes the lowest-order term in an 
expansion in orders of $1 / \omega$ (Eq. \ref{magnus}).
In principle, higher-order terms could also be derived. However, in the
limit $1 / \omega \rightarrow 0$ in which we work, their influence is small,
and can be rendered negligible by raising $\omega$ sufficiently. 
It is straightforward to calculate the true quasienergies of
the driven BH model numerically by diagonalizing the time-evolution
operator for one period, $U(T,0)$ 
\begin{equation}
U(T,0) =  {\cal T} \left( \exp\left[ -i\int_0^T  H(t') dt') \right] \right) 
\label{evolve}
\end{equation}
where $\cal T$ is the time-ordering operator.
The eigenvalues of $U(T,0)$, $\{ \lambda_j \}$, are simply
related to the quasienergies via 
$\lambda_j = \exp \left[-i T \epsilon_j \right]$.
In Fig. \ref{energy_scaling} we compare the energy levels of $\Heff$
with the true quasienergies as the driving frequency is increased. 
For the two cases shown, it can be clearly seen
that for $\omega / U > 50$, the eigenvalues of $\Heff$ coincide with the
Floquet quasienergies to an accuracy of better than $0.1 \%$. Furthermore, the 
differences between the energy levels and quasienergies scale quite accurately 
as $\left( U / \omega \right)^2$.
We have further verified that the same scaling occurs for the complete 
energy spectrum over the entire range of $\kappa$. This confirms that $\Heff$
is indeed an accurate and well-controlled approximation to the full Floquet
operator in both the Mott insulator and superfluid regimes, and that $\Heff$
becomes exact in the limit $\omega \rightarrow \infty$.

In contrast to the case of potential driving, we see that there are
no single-particle hopping terms in this effective Hamiltonian. 
Instead, the dynamics arises from four-operator terms, such as the motion 
of doublons and assisted tunneling processes \cite{assisted}.
As a consequence the motion of particles is highly correlated, producing
a novel type of flat-band system. 
When the driving parameter
$\kappa$ is set to zero, the ground state of the model for commensurate
filling is again
a Mott insulator \cite{njp_2018}. As $\kappa$ is increased, the system makes
a transition to an exotic superfluid state \cite{prr_2019,njp_2023}. 
Just as with the conventional BH model, we will label these two regimes of the kinetically-driven BH model by the
behaviour of the ground-state of the commensurate system, so that
low-$\kappa$ corresponds to the Mott regime, and high-$\kappa$ to the superfluid.
As shown in 
Fig. \ref{phase_transition}b, in the superfluid regime
the momentum density shows two peaks, indicating the macroscopic occupation of
two momentum states, $k = \pm \pi/2$. Reference \cite{prr_2019} 
showed that the ground state is actually a cat-like superposition
of these two many-particle momentum states, and uncovered the
particular importance of the two-particle scattering processes
$\left( k, \pi - k \right) \rightarrow \left( k', \pi - k' \right)$, 
which energetically favor the occupation of these momenta.

\begin{table}
        \begin{center}
                \begin{tabular}{c c c c }
                        \hline\hline
                        \addlinespace[1.2ex]
                        $(N, L)$ & $\text{dim}(\mathscr{H})$ & $\text{dim}(\mathscr{H}_0)$ & $\text{dim}(\mathscr{H}^+_0)$ \\
                        \addlinespace[1.2ex]
                        \hline
                        \addlinespace[1.2ex]
                        $(7,8)$ & $3432$ & $429$ & $232$ \\
                        $(8,8)$ & $6435$ & $810$ & $440$ \\
                        $(9,8)$ & $11440$ & $1430$ & $750$ \\
                        $(8,9)$ & $12870$ & $1430$ & $750$ \\
                        $(9, 9)$ & $24310$ & $2740$ & $1387$ \\
                        $(10,11)$ & $184756$ & $16796$ & $8524$ \\
                        $(11, 11)$ & $352716$ & $32066$ & $16159$ \\
                        $(11, 12)$ & $705432$ & $58786$ & $29624$ \\
                        $(12, 12)$ & $1352078$ & $112720$ & $56822$ \\
                        \addlinespace[1.2ex]
                        \hline\hline
                \end{tabular}
        \end{center}
        \caption{Dimension of the Hilbert space $\mathscr{H}$ and its subspaces.
We label the system by $(N, L)$ where $N$ is the number of bosons and
$L$ is the number of lattice sites. 
$\mathscr{H}_0$ denotes the subspace
of zero-momentum states, while $\mathscr{H}_0^+$ signifies the
subspace of states with both zero momentum and reflection symmetry.
The dimension of $\mathscr{H}$ is simply given by the
binomial coefficient $\text{dim}(\mathscr{H}) = \binom{N+L-1}{N}$;
the other dimensions are obtained numerically by counting states.}
 \label{table:dimension}
\end{table}

\section{Results}
\label{results}
To obtain the energy levels of the Hamiltonians (\ref{eq:Hcbh}) and (\ref{eq:Hkdbh}),
we employ an exact diagonalization method, working in a momentum-space
basis. In all cases we apply periodic boundary conditions.
The Hamiltonians possess a number of discrete symmetries such
as momentum conservation and parity symmetries, and in order
to investigate the spectral statistics of the system it is necessary
to study each subspace separately. Following Ref. \cite{kollath_2010}
we work in the subspace of states with translation
invariance, $a_j \to a_{j+1}$ in real space,
which is equivalent to considering states of zero total momentum,
and reflection symmetry $a_{k} \to a_{-k}$ 
in momentum space. In Table \ref{table:dimension} we give the dimensions
of the Hilbert spaces and subspaces obtained by imposing these symmetries
for systems of $N$ bosons in an $L$-site lattice, which we shall denote
as $(N,L)$. Note that imposing these two symmetries considerably reduces the
size of the Hilbert space,
allowing the diagonalization of systems as large as $(12,12)$ on a
standard desktop computer.

\subsection{Energy level spacings}
The most direct way to study the spectral statistics of a system is to look
at the spacings between consecutive energy levels, $s_n = E_{n+1} - E_n$.
Following the Berry-Bohigas conjectures, for an integrable system the probability 
that a normalized spacing lies between $s$ and $s + ds$ should be given
by the Poisson distribution
\begin{equation}
P(s) = e^{-s} \ ,
\label{poisson}
\end{equation}
while in contrast, chaotic systems should exhibit GOE statistics
\begin{equation}
P(s) = \frac{\pi s}{2} e^{-\pi s^2/4} \ .
\label{goe_statistics}
\end{equation}

The level spacings will have a smoothly varying component depending on the 
local density of states, and
fluctuations around this smooth behavior. It is these fluctuations which
are expected to have the universal behaviour described by Poisson statistics
or RMT. It is thus necessary to first remove the smooth part,
a procedure which is known as ``unfolding''. For some systems, such
as billiards, the smooth behavior is known analytically, but in general
the choice of smooth function is somewhat arbitrary and can 
influence \cite{misleading} the results. 

\begin{figure}
\begin{center}
\includegraphics[width=0.45\textwidth,clip=true]{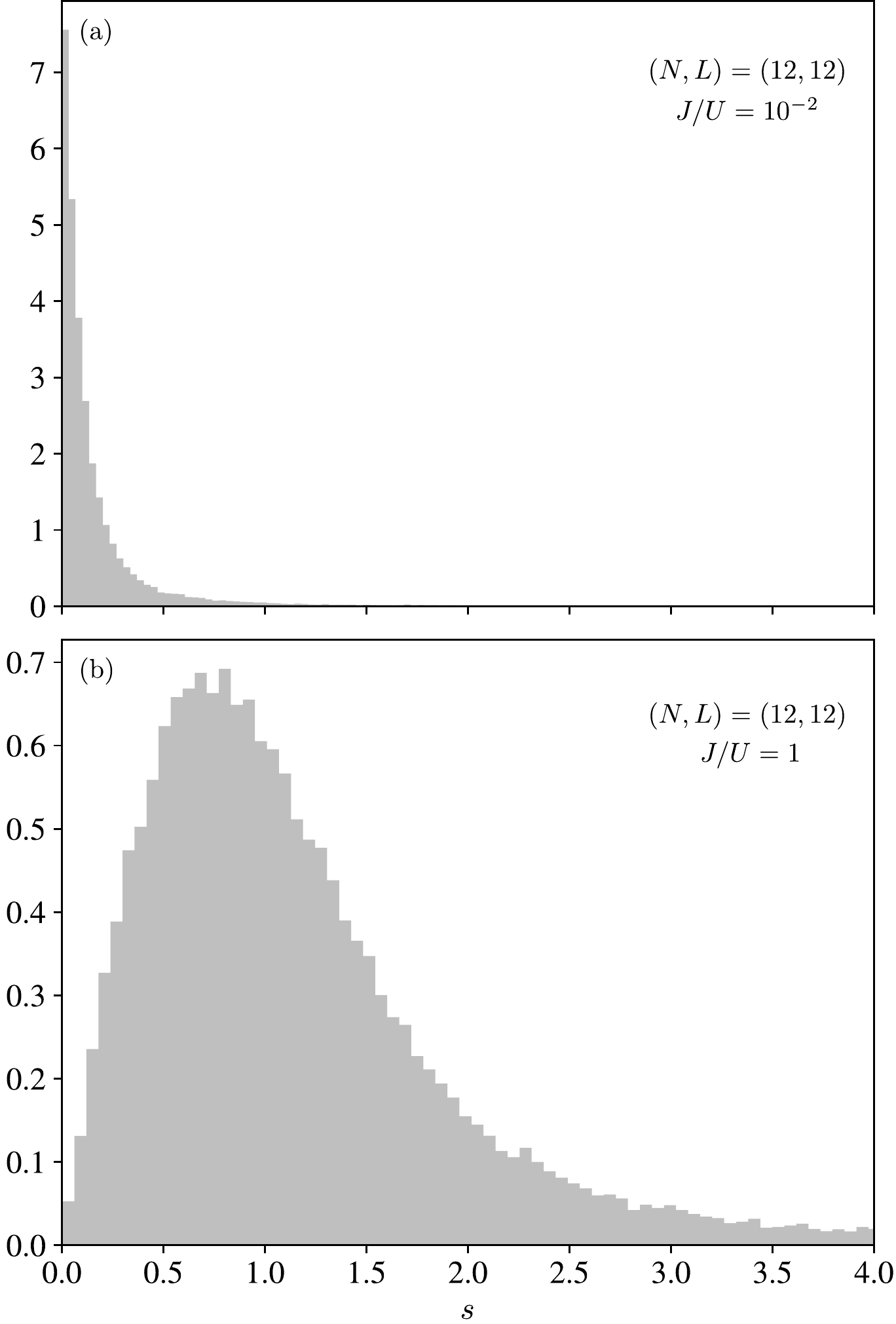}
\end{center}
\caption{Histograms of normalized level spacings for the conventional BH model.
(a) For $J / U = 10^{-2}$ the system is in the Mott regime,
and the spacings show a clear Poisson distribution (Eq. \ref{poisson}).
(b) For $J / U = 1$ the system is in the superfluid regime,
and the spectral statistics change to GOE form (Eq. \ref{goe_statistics}).
In particular, note the suppression of spacings for $s \simeq 0$,
indicating the presence of avoided crossings and thus quantum chaos.
In both cases the system size is $(12,12)$.}
\label{spacings_CBH}
\end{figure}

In Fig. \ref{spacings_CBH} we show the histograms obtained for the conventional BH model
for a large $(12,12)$ system at a low and a high value of $J / U$.
In both cases we used a fitted cubic polynomial to perform the unfolding.
In the limit $J / U \to 0$, the system will be in the Mott
regime and so will be trivially integrable. Accordingly we see in
Fig. \ref{spacings_CBH}a that the level spacings
histogram for the lower value of $J / U = 0.01$ has a strikingly
Poissonian form, peaked at zero with an exponential decay.
As $J / U$ is increased, the system will leave the Mott regime
and enter the more weakly-interacting superfluid regime.
In this regime the distribution of level spacings shown in  
Fig. \ref{spacings_CBH}b has GOE form. In particular we can note
the suppression of the distribution for $s \simeq 0$, indicating the
development of avoided crossings in the spectrum, and thus the
presence of quantum chaos. Similar results are also obtained for
the $(11,12)$ case (not shown). 

\begin{figure*}
\begin{center}
\includegraphics[width=0.9\textwidth,clip=true]{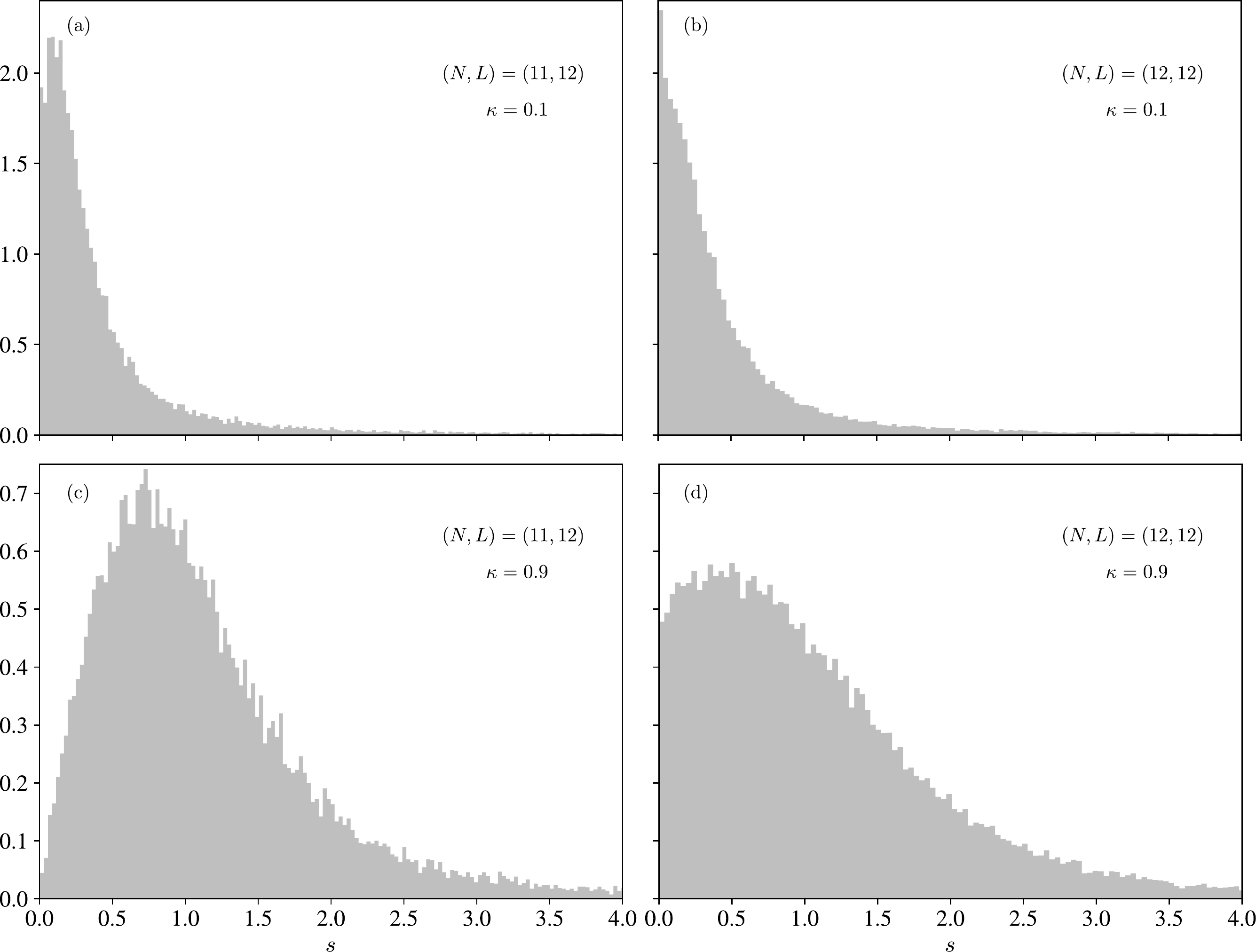}
\end{center}
\caption{Histograms of normalized level spacings  for the kinetically-driven BH model.
(a) For $\kappa = 0.1$ the $(11,12)$ system is in the Mott regime,
and its energy spacings are Poisson distributed.
(b) Similarly, the $(12,12)$ system is also Poissonian for $\kappa = 0.1$.
(c) For $\kappa = 0.9$, the $(11,12)$
system is in the superfluid regime and the statistics become GOE.
(d) Unlike in (c), the $(12,12)$ system does not
exhibit GOE statistics in the superfluid regime. Instead the distribution
appears to be intermediate between Poisson and GOE.}
\label{spacings_KDBH}
\end{figure*}

In Fig. \ref{spacings_KDBH} we show the equivalent histograms for the
kinetically-driven BH model. For a low value of the driving parameter,
$\kappa = 0.1$ the system will be in the Mott regime,
and just as before the histograms indeed show a Poisson distribution
(Figs. \ref{spacings_KDBH}a and b).
Like $H_{\mathrm{BH}}$, $\Heff$ is also a hermitian operator with
time-reversal symmetry, and so we expect its spectral statistics
also to be described by the GOE 
\footnote{As $\Heff$ is obtained by a Floquet method, we must
also ensure that $\omega$ is sufficiently large for all the quasienergies
to lie in the first quasienergy Brillouin zone. If not, some quasienergies
will wrap around the Brillouin zone, and the system 
will instead be described by the circular orthogonal ensemble. As we work in 
the high-frequency regime, $\omega \gg U$, this condition is amply satisfied.}
away from the integrable limit.
In Fig. \ref{spacings_KDBH}c we see that
for the case of $(11,12)$ (11 bosons on a 12-site lattice) the distribution
indeed changes to GOE form as $\kappa$ is increased, 
and the system passes from the Mott to the superfluid regime.
For the $(12,12)$ case, however, the behaviour for
large $\kappa$ is less clear (Fig. \ref{spacings_KDBH}d).
While the distribution is not Poissonian, it shows less 
suppression for $s \simeq 0$ than the GOE distribution, and
appears to be intermediate  between these two distributions.
To investigate this behaviour, and to obtain more quantitative
estimates for the spectral distributions, we will now proceed to analyze
the statistics of the  gap-ratios. 

\subsection{Gap ratios}
The ratio of consecutive gaps is defined as
\begin{equation}
r_n = 
\mathrm{min} \left( \frac{s_{n+1}}{s_{n}}, \frac{s_{n}}{s_{n+1}}  \right) \ .
\label{r_parameter}
\end{equation}
This quantity was introduced by Oganesyan and Huse \cite{oganesyan_2007},
and has the advantage that its measurement does not require the unfolding
procedure, as the ratio of consecutive
spacings does not depend on the local density of states. 
The statistical distribution of $r$ is straightforward to obtain
for the Poisson case
\begin{equation}
P(r) = \frac{2}{\left( 1 + r \right)^2} \ ,
\label{r_poisson}
\end{equation}
and Wigner-like surmises have been obtained \cite{atas_2013,armando_surmise} 
for the RMT ensembles.

\begin{figure}
\begin{center}
\includegraphics[width=0.45\textwidth,clip=true]{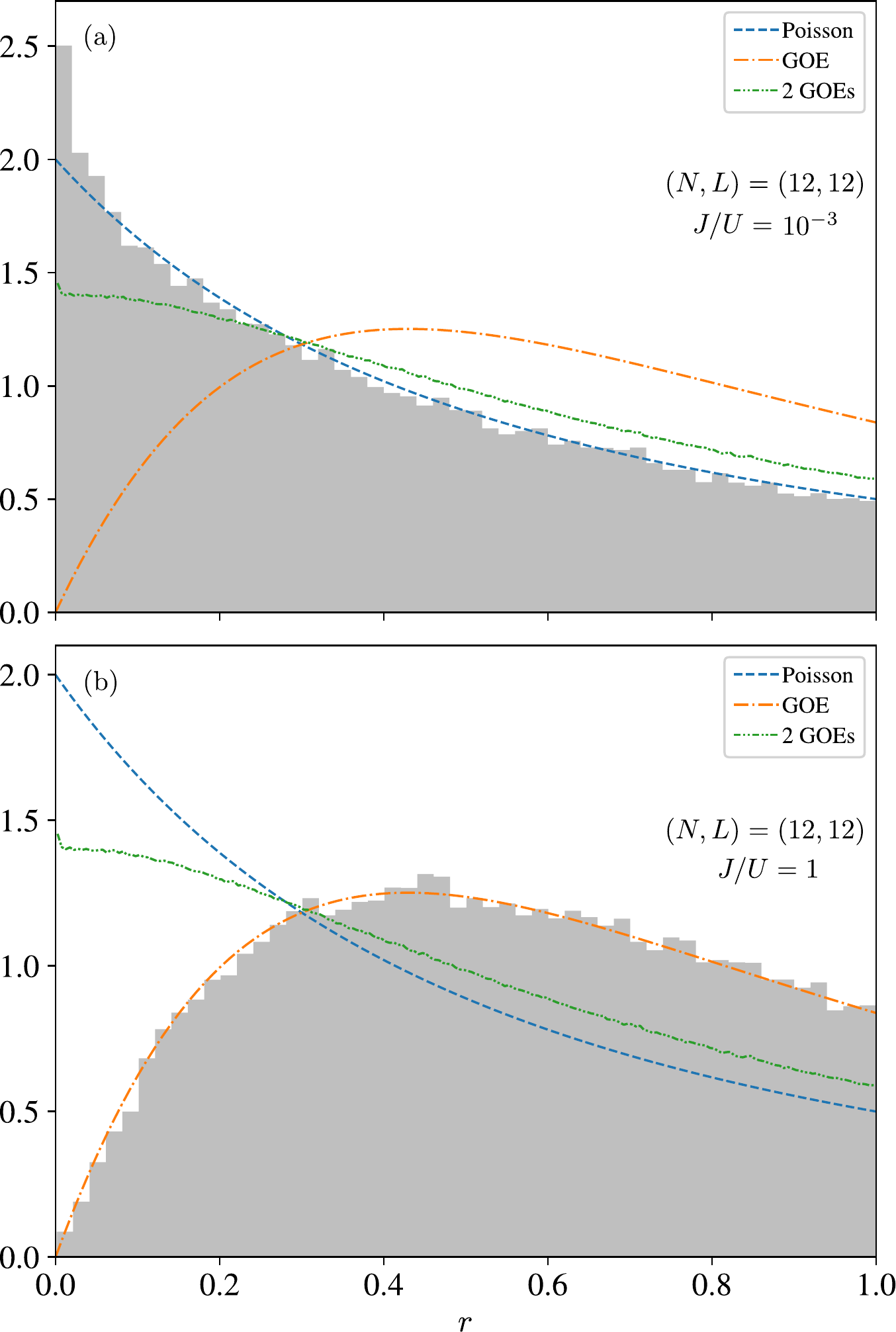}
\end{center}
\caption{Histograms of the gap-ratios for the conventional BH model.
(a) In the Mott insulator regime, $J / U = 10^{-3}$,
the gap-ratios arise from a Poissonian distribution of
level spacings (Eq. \ref{r_poisson}).
(b) For $J / U = 1$, in the superfluid regime,
the distribution of gap-ratios is well-described by the GOE,
in agreement with the previous results for the level spacings.
In both cases the system size is $(12,12)$.}
\label{r_CBH}
\end{figure}

We show results for the conventional BH model in Fig. \ref{r_CBH}. The results
corroborate the behaviour seen previously in the gap
spacings: the Mott insulating system obtained for low $J / U$
exhibits clear Poissonian statistics, while the superfluid case,
obtained for $J / U = 1$, exhibits GOE statistics.
The excellent agreement between the numerics and the analytical
results is quite striking, and underlines the power of using
spectral analysis to discriminate between integrable and quantum chaotic systems.

\begin{figure*}
\begin{center}
\includegraphics[width=0.9\textwidth,clip=true]{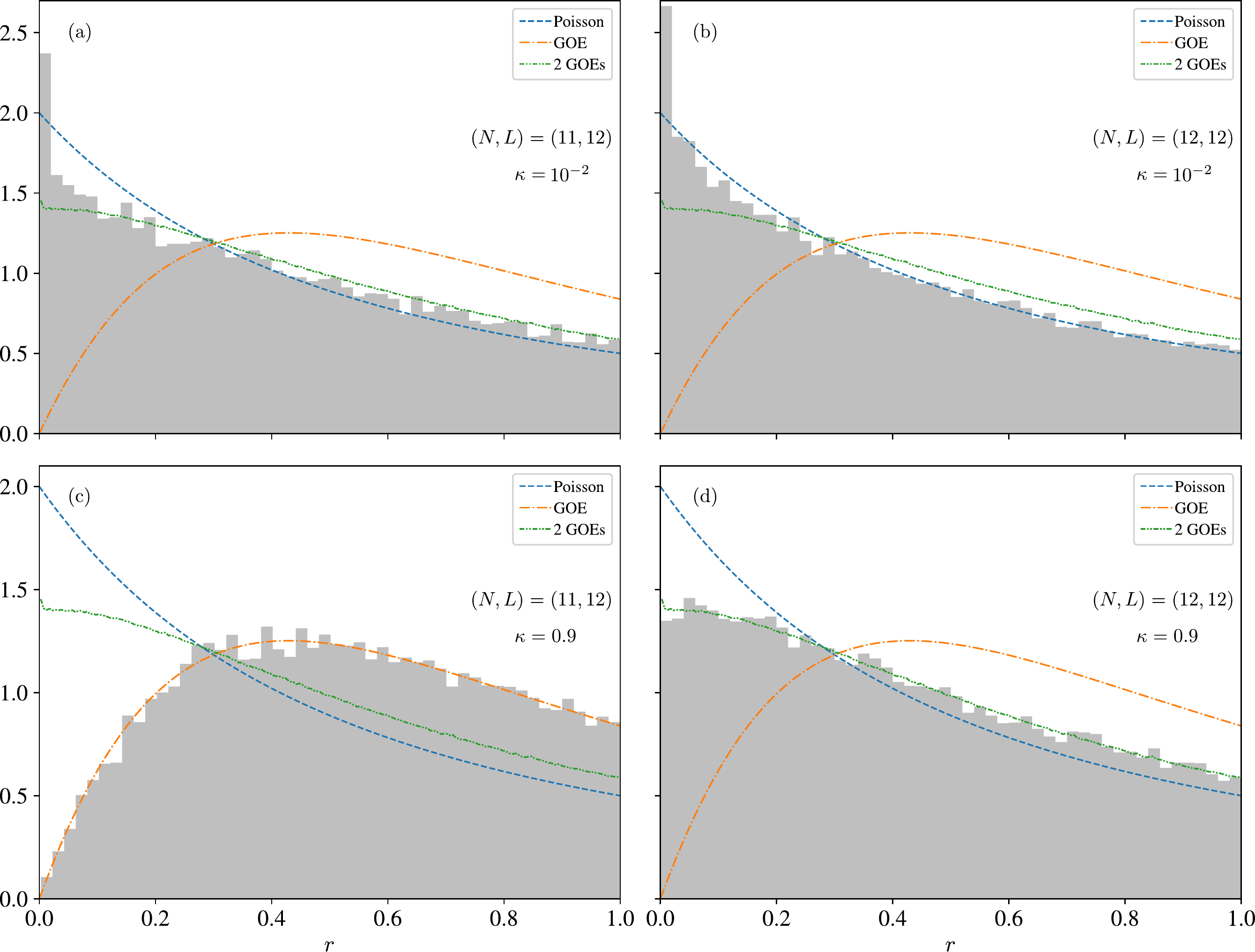}
\end{center}
\caption{Histograms of the gap ratios for the kinetically-driven BH model.
(a) For $\kappa = 0.01$ the $(11,12)$ system is in the Mott regime
and the distribution of gap-ratios is well-described by the
Poisson result. 
(b) Similarly, the $(12,12)$ system is also Poissonian for $\kappa = 0.1$. 
(c) For $\kappa = 0.9$, the $(11,12)$
system makes a transition to the superfluid regime and the statistics 
of the gap-ratios is again GOE.
(d) Unlike in (c), the $(12,12)$ system does not
follow simple GOE statistics in the superfluid regime. Instead the distribution
is described accurately by a mixture of two GOEs.}
\label{r_KDBH}
\end{figure*}

In Fig. \ref{r_KDBH} we show the analogous results for the kinetically-driven BH model
for the two system sizes considered previously. Similarly to the case
of the conventional BH model, the distribution of gap-ratios for the $(11,12)$
system (Fig. \ref{r_KDBH}a and c) changes from
Poisson to GOE statistics as the system passes from the Mott regime
to the superfluid. Again, however, the $(12,12)$ 
(Fig. \ref{r_KDBH}b and d) exhibits an unusual behaviour: in the strongly-correlated
regime the gap-ratios have a Poissonian distribution as expected, but
in the superfluid regime the distribution is neither Poisson nor GOE.
Rather, as we will discuss later, it seems to fit a distribution of
two mixed GOEs.

\begin{figure*}
\begin{center}
\includegraphics[width=0.95\textwidth,clip=true]{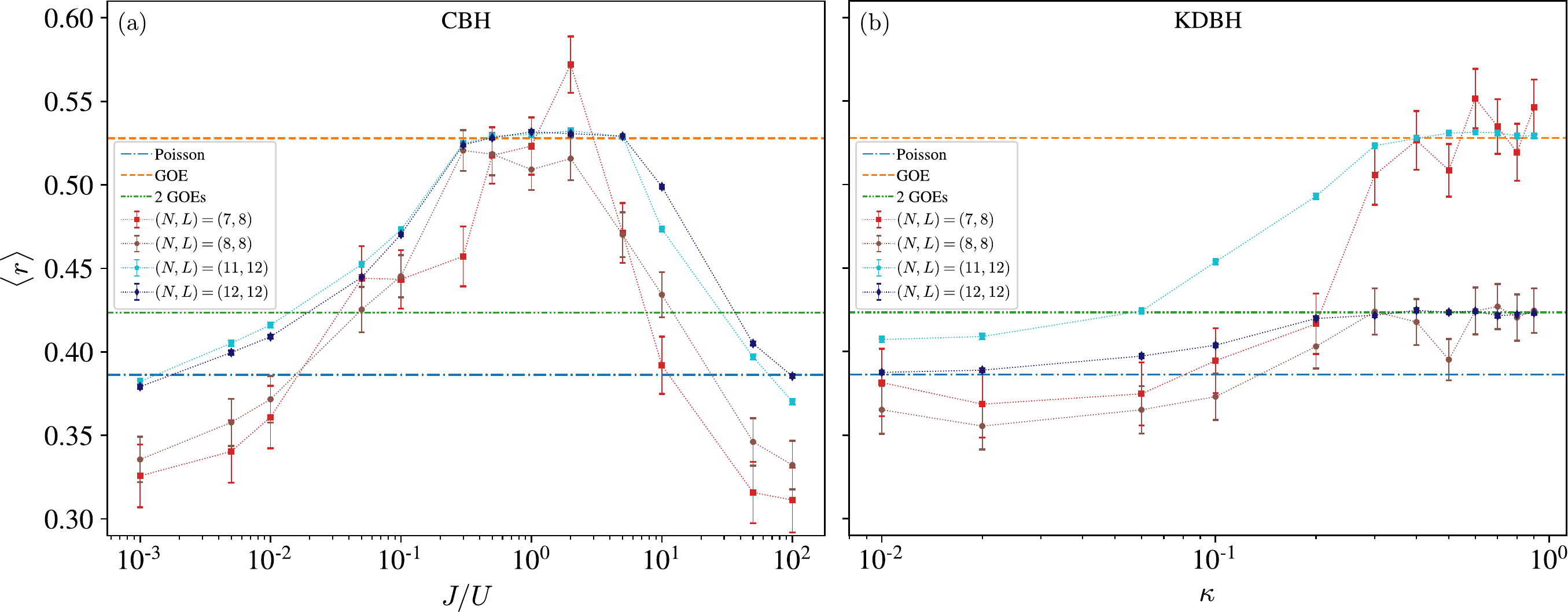}
\end{center}
\caption{Mean gap-ratio $\aver$ as a function of the control parameter.
(a) Conventional Bose-Hubbard model (CBH). This model admits two integrable limits:
$J/ U \to 0$ when the system tends to a Mott insulator, and
$J/U \to \infty$ when it describes a gas of free bosons. In both
cases $\aver$ indicates that the spectral statistics are Poissonian.
Between these limits, in the superfluid regime,
the statistics are GOE. Strong dependence on particle number or
system size is not evident. 
(b) Kinetically-driven BH (KDBH). For small $\kappa$, in the Mott regime
all the system sizes have a Poissonian distribution. In the 
superfluid regime, however, two different behaviors are evident.
For $(7,8)$ and $(11,12)$ the statistics becomes GOE, just as seen
in the conventional BH model. For $(8,8)$ and $(12,12)$, $\aver$ tends
to a value characteristic of a mixture of two GOEs, $\aver = 0.423$, indicating
that these systems have an additional internal symmetry.}
\label{mean_r}
\end{figure*}

From their statistical distributions it is straightforward to calculate the
mean gap-spacing, $\aver$, which can be used to quantitatively assess
to which statistical ensemble a given spectrum belongs.
For the Poisson distribution, $\aver_{\mathrm{Poi}} = 2 \log 2 - 1 \simeq 0.386$,
while for the GOE case $\aver_{\mathrm{GOE}} \simeq 0.528$
\footnote{This quantity cannot be calculated analytically, and
the various surmises provide different estimates of
its value, which, however, only 
differ by a few percent. In this paper we make
use of the values given in Ref. \cite{armando_surmise}.}.
In Fig. \ref{mean_r} we plot the values of $\aver$ for the conventional and kinetically-driven BH models
for several different choices of boson-number and system size, as
we vary the driving parameters. In each case
the size of the error bars was estimated by making a bootstrap
over 4,000 resamples.

The case of the conventional BH model is straightforward
to interpret. This model is integrable in two extremes; for the Mott
limit, as we have already seen, when $J / U \to 0$, and also in the
limit $J / U \to \infty$ when the system becomes a free boson gas.
Accordingly we can see in  Fig. \ref{mean_r}a that $\aver$ takes a value
close to that arising from Poissonian statistics at these limits.
As was previously seen in Ref. \cite{kollath_2010},
some variation from the Poissonian value of $\aver$ is visible, 
especially for the smaller system sizes. In the Mott limit this is a
consequence of the difficulty of calculating the statistics
of a spectrum consisting of well-separated  clusters of nearly degenerate energy levels.
For intermediate values of $J / U$ the value of $\aver$ saturates to a value
close to the GOE average, indicating that the conventional superfluid
exhibits quantum chaos \cite{kolovsky_2004,kollath_2010}.
We observe only minor dependence on the system size,
with the width of the GOE region slightly broadening as the system size increases,
but no dependence on the parities of $N$ and $L$.

Integrability is a fragile property, and so one might expect the spectral 
statistics to abruptly change from Poisson to GOE as soon as the system 
is tuned away from an integrable point. Instead Fig. \ref{mean_r}a shows
a rather smooth crossover between the limits. Whether the GOE region
continues to broaden, thereby making the transition become
sharper as the system approaches the thermodynamic limit remains an open question,
and would require the study of even larger systems. As we can see from 
Table \ref{table:dimension}, however, the rapid growth in the size of the
Hilbert space as $N$ and $L$ are increased make addressing this question
extremely challenging.

As we have seen previously, however, the results for the kinetically-driven BH model are more
difficult to interpret. Unlike the conventional BH model, the system only passes through
a single phase transition as the driving parameter is varied.
For small $\kappa$ all the system sizes considered exhibit Poisson statistics,
corresponding to the integrability of the model in this limit. 
As $\kappa$ is increased, we again see that $\aver$ smoothly increases away
from this value, similarly to the behaviour seen in the conventional BH model.
At a critical value of $\kappa$
the system makes a transition to the superfluid state, but unlike in
the conventional BH case, we do observe a clear dependence on the parities of
$N$ and $L$ of $\aver$ in this regime. When
both the particle number and the number lattice sites are
even, $\aver$
approaches a value intermediate between the Poisson and GOE values, and
otherwise $\aver$ approaches the GOE value.

\begin{figure}
\begin{center}
\includegraphics[width=0.45\textwidth,clip=true]{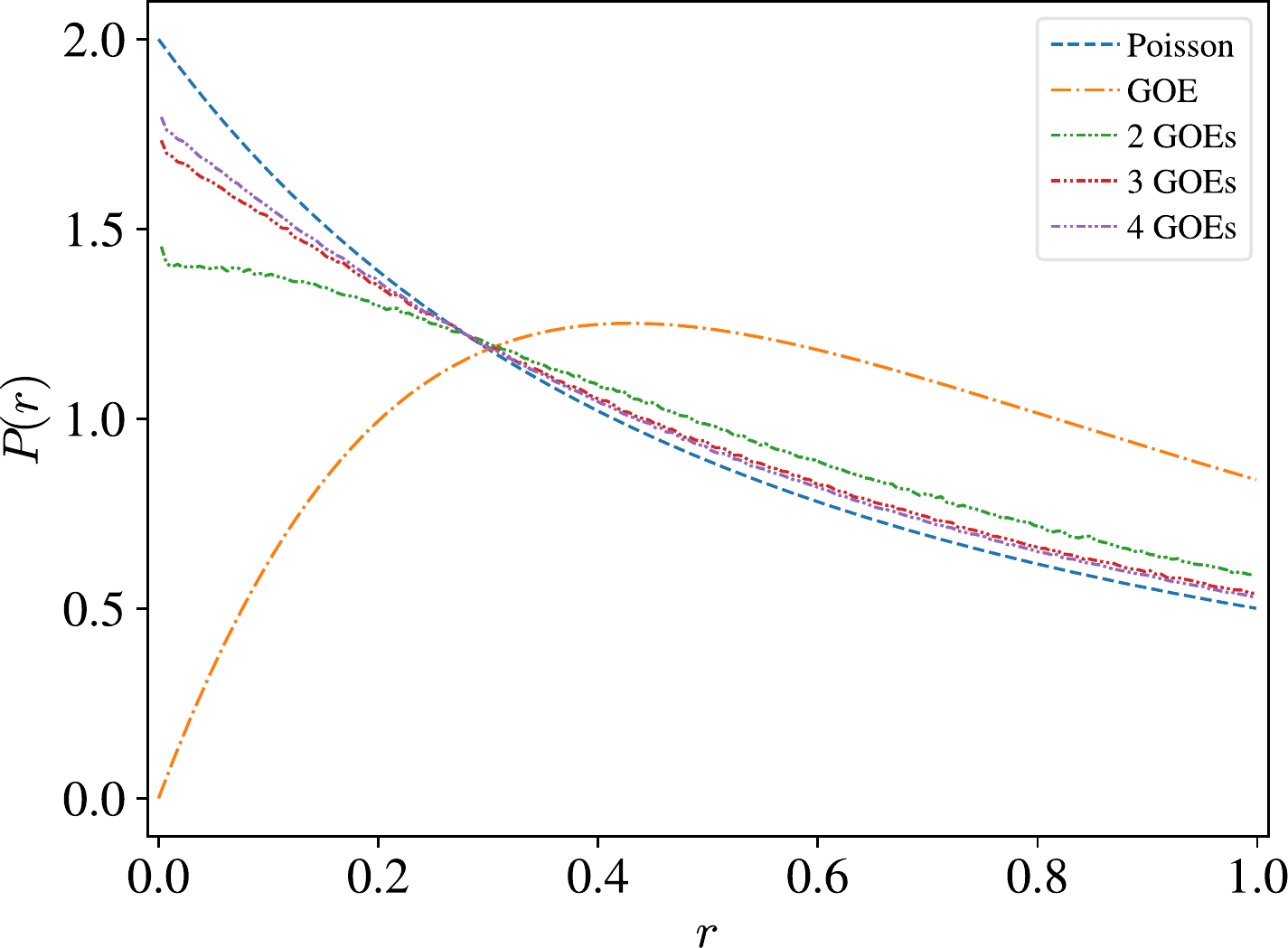}
\end{center}
\caption{Distribution of gap-ratios for several different cases: Poisson
and GOE, and then from the mixture of two, three and four GOE matrices.
The GOE distribution shows the characteristic suppression for small $r$, indicating
the presence of quantum chaos. The mixed results  do not show this feature, and
rapidly approach the Poisson result as the number of independent GOE matrices is increased.}
\label{mixed_blocks}
\end{figure}

Obtaining statistics intermediate between GOE and Poisson is characteristic
of systems containing a hidden symmetry. In principle identifying this
symmetry would allow the Hamiltonian to be divided into independent blocks,
each of which would individually be described by RMT. 
Refs. \cite{rosenzweig,giraud_2022} calculated the statistics for systems
composed of several sub-blocks, and we show the gap-ratio distributions
for several examples in Fig. \ref{mixed_blocks}.
For a single GOE, the gap-ratios show the characteristic
suppression for small $r$, arising from level repulsion, indicating the
presence of quantum chaos. For the case of two mixed GOEs, however,
this feature vanishes, and the resulting distribution is more
similar to, but clearly distinct from, a Poisson distribution \cite{poilblanc_1993}. 
This tendency continues as the number
of mixed GOEs increases, and indeed as the number of mixed spectra
approaches infinity, the distribution tends towards Poisson \cite{giraud_2022}.

The value obtained for $\aver$ for the $(8,8)$ and $(12,12)$ agrees well
with that obtained for two mixed GOEs \cite{giraud_2022}, $\aver = 0.423$,
as can be seen from Fig. \ref{mean_r}b.
To verify this further, the statistical distribution of the gap-ratios
is also plotted in Fig. \ref{r_KDBH}, and again the agreement is seen to be excellent,
particularly in the high-$\kappa$ limit.
Thus for the kinetically-driven BH model we see two types of spectral statistics in the
superfluid regime: a single GOE in general, but two
mixed GOEs when $N$ and $L$ are both even.
Additional examples for different values of $N$ and $L$
are given in the Appendix.

\subsection{Hidden symmetry}
\label{symmetry}
The study of the gap-ratios clearly indicates that the spectral statistics
of the kinetically-driven BH model when both $N$ and $L$ are even are described by a mixture
of two GOEs in the superfluid phase. This points to the presence of
an additional symmetry in the kinetically-driven BH Hamiltonian beyond the translation
and reflection symmetries already accounted for, which depends on the
parities of $N$ and $L$. Furthermore this symmetry must
arise from the kinetic driving, as it is not present in the undriven
BH model.

As noted previously, an essential property of the kinetically-driven BH model is the
importance of pairing correlations between momenta $\pi$ and $\pi - k$. If we define
the $\pi-$reflection transformation, $\mathcal{R}_\pi$, as
\begin{align}
        \label{eq:Q_trans}
        \mathcal{R}_\pi: a_k \to a_{\pi - k} \ .
\end{align}
it is straightforward to show that the kinetically-driven BH Hamiltonian is {\em exactly} invariant
under this transformation for all $k$, while the conventional BH model is not. In addition, under
this transformation, the total momentum of the system transforms as
\begin{align}
        \label{eq:Qpi}
        Q\to \tilde Q = \sum_{\ell=0}^{L-1}
       \left(\pi- {k_\ell} \right) n_{k_\ell} = \pi N -Q \ .
\end{align}
From this it is clear that the $\pi-$reflection symmetry is only compatible with the
$Q=0$ momentum sector if $N$ is even. For this reason, if $N$ is odd this
symmetry is absent, and the spectral statistics arise from a single GOE. 
It is also important to note that $\mathcal{R}_\pi$ can only
be defined when $k=\pi$ is an allowed wave vector in the ring.
This places the additional constraint
that $L$ must be even. 
The presence of this symmetry thus precisely explains 
the conditions on GOE mixing observed in Fig. \ref{mean_r}b. 

\begin{figure}
\begin{center}
\includegraphics[width=0.45\textwidth,clip=true]{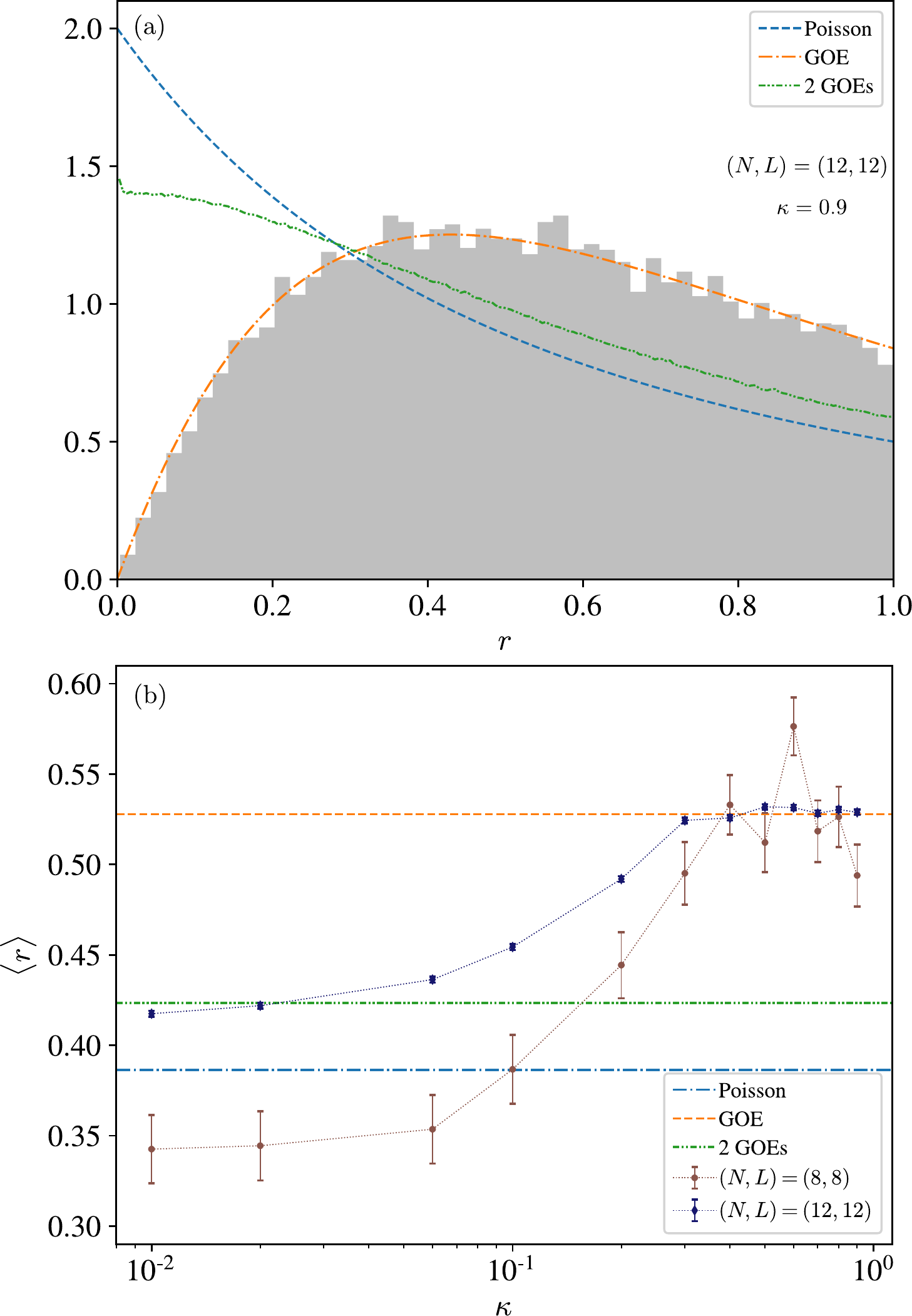}
\end{center}
\caption{Distribution of gap-ratios for the kinetically-driven BH model, applying
the $\pi$-reflection symmetry to the effective Hamiltonian (\ref{eq:Hkdbh}).
(a) Previously (Fig. \ref{r_KDBH}d) the distribution was a mixture of
two GOEs. Applying the symmetry separates the two subspaces, so now
the distribution is described by a single GOE.
(b) Plotting the mean gap-ratio shows that the subspace of states indeed
makes a transition from Poissonian statistics to the statistics of a single GOE
as $\kappa$ is increased.}
\label{r_pi}
\end{figure}

Knowing this symmetry, it is thus possible to symmetrise the kinetically-driven BH Hamiltonian
and divide it into two blocks, each block corresponding to one parity
of $\mathcal{R}_\pi$. We show the results in Fig. \ref{r_pi}. Clearly the 
results show that separating the two subspaces in this way indeed yields a single
GOE distribution. In particular we note that the value of $\aver$ now
tends to the GOE value of 0.528 in the limit of large $\kappa$, exactly
as we would expect. We can also see that the location of the crossover
again shows a weak dependence on $N$, moving to a smaller value of $\kappa$
as the system size is increased. As with the conventional BH model, this behaviour
is broadly consistent with the transition occurring for an infinitesimal
value of $\kappa$ in the thermodynamic limit.
But also like the conventional BH case, the precise behaviour of the Poisson-GOE crossover 
in the thermodynamic limit is an open question.

\section{Conclusions}
In summary, we have studied the spectral statistics of the Bose-Hubbard model
under kinetic driving, and compared and contrasted its behaviour with
that of the conventional BH model. When the driving parameter $\kappa$ is small,
the system lies in the Mott regime, and makes
a transition to a superfluid behaviour as $\kappa$ is increased. We have
seen that, as also occurs in the conventional BH model,
this transition is accompanied by a change in the spectral
statistics: from Poisson to GOE. 
As a similar correlation between the nature of the ground state
and the spectral statistics
has also been seen previously in the conventional BH model \cite{kollath_2010}, it is interesting
to speculate that this somewhat unexpected link may be a general feature seen in systems
tuned away from an integrable point.
This may well be the case, at least for systems with a superfluid ground state, because the same interactions that are needed to create a robust superfluid state \cite{F20,F27} can be expected to yield quantum chaos in the excited states. This correspondence is likely to be maintained in Bose systems with higher dimensions \cite{wimberger}, for which integrability is less protected. 

In contrast to the
case of the conventional BH model, however, 
we have found that the form of the GOE statistics in the
superfluid regime depends on the parities
of the number of sites and the number of particles. 
When both of these quantities are even, evaluating
the gap ratio of the spectrum has allowed us to establish
that the system is accurately described as
the mixture of two GOEs. We have further shown that this arises from
an emergent symmetry we term ``$\pi$-reflection'' invariance, produced
by the kinetic driving. When this symmetry is not present, meaning that
either $N$ or $L$ is odd, the system is described as a single GOE,
characteristic of quantum chaos. We have thus seen how spectral 
statistics can be used as a probe to uncover hidden symmetries
in a many-body quantum system.

Similarly to the case of the conventional BH model, the crossover in the spectral
statistics from Poisson to GOE
in the kinetically-driven BH model shows a weak dependence on the system size,
moving towards smaller values of $\kappa$ as the number of sites is
increased. This behaviour is not inconsistent with the system making
the transition to chaos for an infinitesimal symmetry breaking in
the thermodynamic limit \cite{rabson, santos_1, santos_2, santos_3},
although the existence of a finite threshold
cannot be completely ruled out. A definitive statement would require
a scaling analysis extending to much larger systems, which is computationally 
unfeasible at the current time, but remains an attractive prospect for
future research.

\acknowledgments
This work was supported by the the Spanish MICINN through Grant Nos. 
FIS2017-84368-P and PID2022-139288NB-100, and the Universidad Complutense de Madrid
through Grant No. FEI-EU-19-12. The authors would like to
thank Guillaume Roux for sharing eigenvalue data for
the conventional BH model, and acknowledge stimulating and
valuable conversations with Armando Rela\~no and \'Angel Corps.

\appendix
\section{Symmetry and parity}
\begin{figure*}
\begin{center}
\includegraphics[width=0.9\textwidth,clip=true]{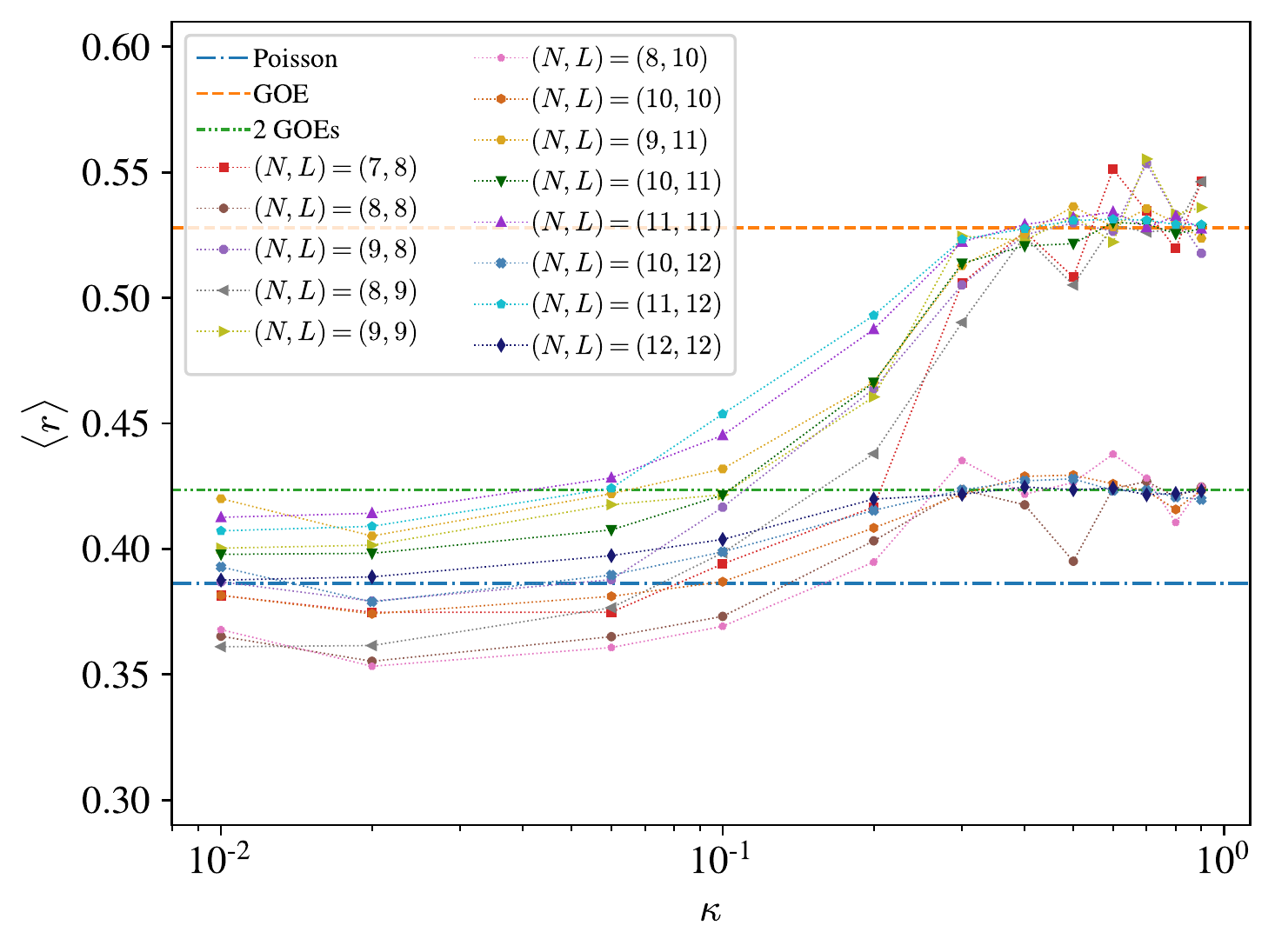}
\end{center}
\caption{Mean gap-ratio, $\aver$ as a function
of $\kappa$ for a variety of choices of $(N,L)$. Two forms of behaviour
are evident. For $N$ and $L$ both even, the distributions tend to a mixture
of two GOEs as $\kappa$ is increased. Conversely, if $N$ or $L$ is
odd, the distribution tends towards a single GOE, in agreement with
the symmetry analysis given in Section \ref{symmetry}.}
\label{summary}
\end{figure*}

For clarity of presentation, 
the results shown in Section \ref{results} were restricted to just a
few representative system sizes, namely $(11,12), \ (12,12), \ (7,8)$ and $(8,8)$.
In Fig. \ref{summary}  we give results for the kinetically-driven BH model for a more exhaustive 
selection of lattice sizes and particle numbers, to illustrate the parity
dependence of the spectral statistics in the superfluid regime.

It is immediately clear that the curves fall into two distinct classes. In all
cases the distributions tend towards being Poissonian in the limit $\kappa \to 0$.
As $\kappa$ is increased, one set of curves evolve toward the GOE result of
$\aver = 0.528$, while the remainder evolve towards the result for
two mixed GOEs. This latter group corresponds to values of $N$ and $L$ which are
both even, while those that evolve  toward a single GOE have one or both of
these quantities being odd.

This corresponds exactly to the symmetry dependence described in Section \ref{symmetry}.
If $L$ is odd, then the momentum $k = \pi$ is not found in the FBZ, and so the
$\pi$-reflection symmetry cannot be implemented. On the other hand, when
$N$ is odd, Eq. \ref{eq:Qpi} is not invariant, and so the symmetry is absent.
As a consequence, $\pi$-reflection symmetry is only realized in systems having
both $N$ and $L$ even.

\bibliographystyle{aipnum4-1}
\bibliography{statistics_bib}

\end{document}